\documentclass{article}
\usepackage[utf8x]{inputenc}
\usepackage{ucs}
\usepackage[OT1]{fontenc}
\usepackage{amsmath}
\usepackage{amsfonts}
\usepackage{amssymb}
\usepackage{authblk}

\usepackage{graphicx}
\usepackage{indentfirst}
\usepackage[left=3cm,right=1cm,top=2cm,bottom=2cm]{geometry}
\begin{document}

 \begin{center}

 \large \bf Velocities of distant objects in General Relativity revisited
  \end{center}

 \vspace{0.3truecm}

 \begin{center}

 E. D. Emtsova$^{1,2}$  and A.V. Toporensky$^{2,3}$

\vspace{0.3truecm}

 \it $^{1}$ Physical Faculty, Lomonosov Moscow State University, Moscow 119991, Russia

 \it $^{2}$ Sternberg Astronomical Institute, Lomonosov Moscow State University,
Universitetsky Prospect, 13, Moscow 119991, Russia

 \it $^{3}$ Faculty of Physics, Higher School of Economics, Moscow, Russia

\end{center}

\begin{abstract}
We consider two most popular definitions of velocities of remote objects in General Relativity.
Our work has two motivations. From a research point of view, we generalize the formula connecting
these two velocities in FRW metrics found by Chodorowski \cite{Hodorowski} to arbitrary synchronous
spherically symmetric metrics. From a methodological point of view, our goal is to outline 
certain counter-intuitive properties of the definitions in question, which would allow to use
them when it is reasonable and to avoid
incorrect statements, based on inappropriate use of intuition.
\end{abstract}


\section{Introduction}

The problem of interpretation of recession velocities has a long story. A seemingly trivial question
often led to mistakes and wrong interpretations, both in scientific and pedagogical literature. A lot of
striking examples have been collected in the paper \cite{Davis}, including incorrect statements
of such prominent researchers as R.Feynman, W. Rindler, S.Weinberg and others. Most part of mistakes
have been connected with the fact that recession velocities can exceed the speed of light, which is
now considered as a well established property following from the non-local nature of
velocities of distant objects. 

However, approach of \cite{Davis, Davis2} is not a unique one. In fact, it used Hubble law to define velocities.
Apart from this definition 
(we will consider it in detail later), there are other proposals, which lead to totally different
properties of recession velocities. In particular, an approach intensively popularized by Synge
in \cite{Sing} never leads to superluminal velocities. This may be considered as an unambiguous
advantage of such definition, and its supporters usually stress this point \cite{Hodorowski}. It was
also stated that, unlike cosmological recession velocities, this approach (which is also explained later
in our paper) can be applied to arbitrary space-times.

We should say, however, that the Synge's definition of velocity has their own features which
can be considered as disadvantages (or, at least, counter-intuitive properties). As for the claimed universality,
the situation is more interesting. In recent methodological paper \cite{we} it was shown that almost
all "cosmological" concepts (namely, those which do not require spatial homogeneity) can work successfully
for static space-times (for example, black hole metrics) if we work in synchronous coordinate system
(the system, connected with a body free falling into a black hole). In particular, Hubble law and recession 
velocities have their analogs in this picture and are not bound with only cosmology.

The goal of the present paper is not to discuss what definition is better. As both are mathematically correct,
they both can be used in calculations. Our goal is to discuss some features of these definitions, which 
can be a source of errors since they look "strange" in comparison with the properties of velocity known
from Special Relativity. In principle there exist a viewpoint (shared by some researchers in extragalactic
astronomy) that it is in general meaningless to think about such immeasurable entities as distance
and velocities of remote galaxies and we should use only measurable object like the redshift. We think, 
however, that it is impossible to forbid  question of how remote distant galaxies are. as well as about
their velocities relative to us
(especially this is important for students studying General Relativity), and it is better to consider correctly
different definitions of these entities with their properties (sometime unexpected). We hope that the
present paper may help to achieve this goal. 

The structure of the paper is as follows: in Sec.2 we remind a reader the definitions of velocity used
in \cite{Davis} and \cite{Sing, Hodorowski}. In Sec.3 we show that  these two velocities 
are connected by rather simple formula in any spherically symmetrical metrics. This result generalizes the relation found for FRW metrics
in \cite{Hodorowski}. In Sec.4 we write down explicit formulae for cosmological and spherically-symmetric
black hole space time, showing counter-intuitive features of the velocities in question (note that for 
different space-times different properties may be considered as "counter-intuitive"). In Sec.5 we summarize
our results in conclusions.

\section{Two different ways for velocities of remote objects}

We begin consideration of the first definition of velocity in a cosmological context, where this
definition appears naturally. The FRW metrics written in the usual form
$$
ds^2=dt^2-a^2(t)(dr^2+r^2 d\Omega^2)
$$
(where $r$ is comoving radial coordinate,
 $d\Omega$ is the angular element) admits  particular foliation of space-time 
by the hypersurfaces $t=const$ such that corresponding spatial slices are homogeneous.
This property is so convenient for this foliation to be used everywhere possible often
without a particular attention. Note, however, that a realistic observer which moves with
respect to the frame of constant $r$ would see inhomogeneous Universe, as it is for Earth-based
astronomers, observing dipole anisotropy of Cosmic Microwave Background. Nevertheless, it is 
much easier to work in the frame with homogeneous space slices, going to other frames only when
it is absolutely necessary. 

Having this foliation it is natural to introduce proper distance to a remote object defined as
$ d = ar$ (assuming that the observer is located at the origin $r=0$) and corresponding velocity
as $v=\dot d = \dot a r$ for an object being in rest with respect to the frame considered. These two
simple formulas give us the Hubble law $v=Hd$ which in the FRW Universe is an exact relation. For
other proposals for FRW cosmology see, for example, \cite{Popov1, Popov2}.

This definition of velocity looks rather natural, however, we immediately can see some unusual 
features. As the Hubble law $v=Hd$ in the FRW metrics is exact, we can apply it to remote object with arbitrary large $d$.
For justify this we need to consider universe without particle horizon. Such models exist
and in these models unboundness of $d$ for objects seen by an observer at $r=0$ ultimately leads to
unboundness of $v$.

If we consider objects with nonzero peculiar velocities, we can see further deviation from 
things familiar from Special Relativity. Indeed, it is reasonable to define
$ v = d(ar)/dt = \dot a r + \dot r a$. This means that if we introduce recession velocity
$v_r=\dot a r$ existing due to nonstationarity of the metrics, and peculiar velocity 
$v_p=\dot r a$ existing due to motion of the object with respect to the FRW frame,
we get $v=v_p+v_r$, the classical Galilean law, independently of the values of velocities.
This leads to relations like $c/2+c/2=c$ (or, even $2c/3+2c/3=4c/3$!) which can shock any person
familiar with Special Relativity.

On the other hand, recession velocities have a "natural" additivity property. Namely, suppose
we have an observer located in $r=0$, first emitter located at $r_1$ and second emitter located
at $r_2$. It is evident from the definition, that the recession velocity of the second emitter
$v_2$ is equal to recession velocity of the first emitter $v_1$ plus the recession velocity $v_{12}$
of the second emitter with respect to the first emitter (since the coordinate $r$ is an additive
variable, and the scale factor is fixed as it corresponds to a fixed time). We will see that such
a "natural" property does not hold for other definitions of velocity.

As it is shown in \cite{we} the definition considered here can be easily generalized to non-cosmological
situations, if we use a synchronous coordinate system. In the absence of homogeneity it is not in general
convenient to put an observer in the origin of the coordinate system, so let the position of an observer
be at $r_1$ and the emitter at $r_2$. 
It is known that
in a sinchronous system coordinate lines are geodesics, so considering a frame in which local observers have
fixed radial coordinate we can define 

\begin{equation}
d=\int_{r _{1}}^{r _{2}}\sqrt{g_{r r }}dr
\end{equation}

and 

\begin{equation}
v=\frac{d}{dt} \int_{r _{1}}^{r _{2}}\sqrt{g_{r r }}dr
=v_{fl}+v_{loc}  \label{sum}
\end{equation}

The velocity naturally decompose in a part originated from the fact that the metric is not 
stationary (an analog of recession velocity of the Hubble flow)
\begin{equation}
v_{fl}=\int_{r _{1}}^{r _{2}}\frac{d}{dt}\sqrt{g_{r r }}dr
\label{bhf}
\end{equation}%
 and a part originated from changes of radial coordinate  (the analog
of peculiar velocity in cosmology)
\begin{equation}
v_{loc}=\sqrt{g_{r r }(r _{2})}dr _{2}/dt  \label{bhl}
\end{equation}

(The above was for the situation when the observer has smaller radial coordinate then the emitter.
This is natural in cosmology, where observer is usually considered to be located in the origin of
the coordinate system. However, in black hole space-times we can meet the
 opposite situation when  the observer is located on the upper limit $r_{2}$, then we define $ v_{loc}=\sqrt{g_{r r }(r _{1})}dr _{1}/dt$ and, so, $v=\frac{d}{dt} \int_{r _{1}}^{r _{2}}\sqrt{g_{r r }}dr
=v_{fl}-v_{loc} $ in order to set $v_{loc}$ to a positive value if it is directed towards increasing
$r$.)

We now turn to other proposal for the velocity which uses completely different ideas. Intuitively, let
us "transfer" the velocity from a distant point to the location of an observer. The parallel transport
on a Riemannian manifold is a well-defined mathematical object. However, we can not apply it directly
to velocities since they are 3-dimensional objects. We can make the parallel transport of 4-vector using appropriate connections
(General Relativity in its standard from uses Levi-Civita connections, we use them in the present paper and will comment
about other choice later), so to start the procedure we take 4-velocity of a distant object and
transport it to the observer point. Then, we restore 3-velocity using transported 4-velocity and
4-velocity of the observer. 

One property of such definition is clear -- any 3-velocity obtained by this procedure is subluminal
(a hypothetical superluminal 3-velocity would correspond to imaginary 4-velocity vector which can
not be a result of parallel transport of any real 4-velocity vector). However, the procedure is still not
fixed completely since for Levi-Civita connection the result of parallel transport depends on the path.
What path is better to specify? One proposal is to chose a null geodesics between the emitter and observer.
This proposal does not require any additional structures, like particular foliation of space-time.
In this sense it can be applied to any space-time. Moreover, 3-velocity defined this way is exactly
the velocity which produces {\it in a flat space-time} the same redshift as the observer sees in curved
space-time -- this rather simple fact can be done to students for an exercise.

(Informal hint: let us express the redshift through the energy ratio of emitted and observed photons
$1+z=(k_{\mu}U^{\mu})_e/(k_{\mu}U^{\mu})_o$. Then transport emitter values to the point of observation.
The scalar product does not change, as for individual meaning of the variables, note that the wave
vector at the point of emission $(k_{\mu}) e$ is transported along null geodesics and thus gives the wave 
vector at the point of observation $(k_{\mu}) o$. As for 4-velocity
of emitter, it gives some transported value $\tilde U^{\mu}$. After that, the standard formula expressing
$z$ through 3-velocity $\tilde V$ can be got exactly the same way as in Special Relativity.)

There are however, arguments against this chose. Usually in physically interesting situations we assume
some foliation by hypersurfaces of constant time. The emitter  sent the light at some time $t_1$
which is earlier than the time when the observer received it $t_2$. This means that the velocity
obtained from parallel transport along the light path has a meaning of an average (in some sense) velocity
in between $t_1$ and $t_2$ (see \cite{Hodorowski}). To construct a velocity {\it at particular time $t$}
we need to transfer 4-velocity along the line $t=const$ -- if we consider only radial motion, the line
of parallel transport
is fully specified. Explicit calculations of such a velocity for FRW Universe have been done in
\cite{Hodorowski} where it was shown that it is connected with the Hubble law velocity (which we 
denoted here as $v_{fl}$) by a simple formula $v=\tanh(v_{fl})$. In the next section we generalize this result 
for any synchronous frame and non-zero peculiar velocities.

Before calculation we would like to point out a conceptual differences between these types of velocities.
If the Hubble law velocity is approximately constant and  is expressed, say, in kilometers per second,this tells us that some distance changes by $v_{fl}$ kilometers during one second (or, very close to this value, if
we decide to be more pedantic). As for the velocity defined via
parallel transport (regardless a particular path used), no physical object covers $v$ kilometers per 1 second.
The 4-velocity of an object being transported along any line different from the world-line of this object
looses any connection with it. Strictly speaking, transported 4-velocity is not a 4-velocity of any
physical object. This leads to some counterintuitive features of transported velocities, which we will see later.

Parallel transport of 4-velocity vector $U^\alpha = \frac{dx^\alpha}{d \lambda}$  along a curve 
 parametrized by   $\lambda$ is defined  by  the differential equation 
\begin{equation} \label{transport}
\cfrac{dU^\alpha}{d \lambda} + \Gamma^\alpha {}_{\beta \gamma} U^\beta \cfrac {dx^\gamma}{d \lambda} = 0
\end{equation}
The initial conditions are the components of the emitter 4-velocity vector.

For Levi-Civita connection the Christoffel symbols are:
\begin{equation} \label{levicon}
\Gamma^\alpha {}_{\beta \gamma} = \frac{1}{2} g^{\alpha \delta} (\partial_{\gamma} g_{\beta \delta}+\partial_{\beta} g_{\gamma \delta}- \partial_{\delta} g_{\beta \gamma})
\end{equation}

We will consider here only radial motion, so $U^2=U^3=0$ and $\theta = const$, $\phi = const$.

For the parallel transport along the constant time curve $t=const$ we have the following. 
Denoting $(x^\mu)=(x^0,x^1,x^2,x^3)$, $x^0=t=const$, $x^2=\theta=const$, $x^3=\phi=const$, the equation \eqref{transport} takes the form:
\begin{equation}\label{teq}
\cfrac{dU^\alpha}{dx^1} = -\Gamma^\alpha {}_{\beta 1} U^\beta
\end{equation}

Let us now consider the parallel transport along the null-geodesic.
Time is increasing along the null-geodesic, so we can write  $x^0 = x^0 (x^1)$ (other  coordinates are constant). Then we rewrite \eqref{transport} : $dU^\alpha = - \Gamma^\alpha {}_{\beta 0} U^\beta dx^0 - \Gamma^\alpha {}_{\beta 1} U^\beta dx^1 = - \Gamma^\alpha {}_{\beta 0} \frac{dx^0}{dx^1} U^\beta dx^1 - \Gamma^\alpha {}_{\beta 1} U^\beta dx^1 $, which gives:
\begin{equation}
\frac{dU^\alpha}{dx^1}  = - \Gamma^\alpha {}_{\beta 0} \cfrac{dx^0}{dx^1} U^\beta  - \Gamma^\alpha {}_{\beta 1} U^\beta 
\end{equation}

Using the condition $dS=0$, we obtain

\begin{equation}\label{svet}
\sqrt{ g_{00}} dx^0 =\pm \sqrt{- g_{11}} dx^1
\end{equation}

This gives 
\begin{equation}\label{seq}
\cfrac{dU^\alpha}{dx^1} = - (\pm \sqrt{-\frac{g_{11}}{g_{00}}} \Gamma^\alpha {}_{\beta 0} + \Gamma^\alpha {}_{\beta 1}) U^\beta
\end{equation}

 We should use "$-$" if the emitter has bigger radial coordinate $r$ than the observer,
"$+$" in the opposite situation

Now we need to recover 3-velocities from transported 4-velocity in particular observation frame.
In the present paper we will consider only observers which are at rest with respect to the
used coordinate frame, so corresponding tetrad, defining the frame is not boosted and not
rotated. 
 In our case this relation is given by:
\begin{equation}\label{VU}
V=\frac{U^{1 '}}{U^{0 '}} = \sqrt{-\frac{g_{11} (t_0,r_0)}{g_{00} (t_0,r_0)}} \frac{U^1}{U^0} 
\end{equation}

where
\begin{equation}
\begin{matrix}
U^{0 '} =\sqrt{g_{00} (x_0)} U^0 =\cfrac{1}{\sqrt{1-V^2}}
\\
U^{1 '} =\sqrt{-g_{11} (x_0)} U^1 =\cfrac{V}{\sqrt{1-V^2}}
\end{matrix}
\end{equation}
- are the tetrad components of 4-velocity vector.

\section{Connecting velocities}

We remind a reader that several years ago Chodorowski have shown that in FRW cosmological metrics two
velocities defined in the previous section are connected by very simple formula, if we use 
parallel transport along $t=const$ line. The goal of this section is to show that this relation
is still valid in any spherically symmetric metrics if we use a synchronous coordinate system.

In the spherically symmetric case and pure radial motion the 4-velocity can be expressed in the
parametric form which we will use in this section.
Namely, using the condition $U_\mu U^\mu = 1$ which now gives: $g_{00} (U^0)^2+g_{11}(U^1)^2=|g_{00}| (U^0)^2-|g_{11}|(U^1)^2=1$, one can write: 
\begin{equation}
\begin{cases}
\sqrt{g_{00}} U^0=\cosh (\alpha)
\\
\sqrt{|g_{11}|} U^1=\sinh (\alpha)
\end{cases}
\end{equation}
So, the 4-velocity vector depends now on one parameter $\alpha$, which changes in some way along the curve of 4-velocity transport.  $\sqrt{g_{00}} U^0 =U^{0 '}$ and $\sqrt{|g_{11}|} U^1 = U^{1 '} $ - are the tetrad components of 4-velocity. Therefore, the 3-velocity in the reference frame in question is $ V =  \tanh  (\alpha) <1 $ and never exceeds the speed of light. 

If the emitter is at rest with respect to the coordinate system used, the parameter $\alpha$ at the beginning
of the parallel transport path (we will use the subscript $(_{*})$ to mark initial values) $\alpha_{*}=0$. For an emitter with non-zero peculiar velocity $v_{loc}$ we
easily get from the definition of $\alpha$ that the initial value is equal to $\alpha_{*}= \operatorname{artanh} v_{loc}$.
During the parallel transport the  vector $ (U ^ {0 '}, U ^ {1'}) $ undergoes a hyperbolic rotation. This rotation can be written as:
\begin{equation} \label{hrot}
\begin{pmatrix}
U^{0 '} \\ U^{1 '}
\end{pmatrix}  = \begin{pmatrix}
\cosh (\Delta \alpha) & \sinh (\Delta \alpha)
\\
\sinh (\Delta \alpha) & \cosh (\Delta \alpha)
\end{pmatrix}  \begin{pmatrix}
U_{*}^{0 '} \\ U_{*}^{1 '}
\end{pmatrix} = \begin{pmatrix}
\cosh (\Delta \alpha) & \sinh (\Delta \alpha)
\\
\sinh (\Delta \alpha) & \cosh (\Delta \alpha)
\end{pmatrix}  \begin{pmatrix}
\cosh (\alpha_{*}) \\ \sinh (\alpha_{*})
\end{pmatrix} = \begin{pmatrix}
\cosh (\alpha_{*} + \Delta \alpha) \\ \sinh (\alpha_{*} + \Delta \alpha)
\end{pmatrix}
\end{equation} 

In order to relate $\alpha$ with the connection coefficients we need the infinitesimal form of \eqref{hrot}.
Infinitesimal rotation on the angle $\delta \alpha$  can be written as:
\begin{equation}
\begin{pmatrix}
U^{0 '} \\ U^{1 '}
\end{pmatrix} + \delta \begin{pmatrix}
U^{0 '} \\ U^{1 '}
\end{pmatrix} = \begin{pmatrix}
1 & \delta \alpha
\\
\delta \alpha & 1
\end{pmatrix}  \begin{pmatrix}
U^{0 '} \\ U^{1 '}
\end{pmatrix} = \begin{pmatrix}
\cosh (\alpha + \delta \alpha) \\ \sinh (\alpha + \delta \alpha)
\end{pmatrix}
\end{equation}

or
\begin{equation}
\begin{cases}
\delta U^{0 '}= U^{1 '} \delta \alpha = U^{1 '} \cfrac{d \alpha}{d \lambda} \delta \lambda
\\
\delta U^{1 '}= U^{0 '} \delta \alpha = U^{0 '} \cfrac{d \alpha}{d \lambda} \delta \lambda
\end{cases}
\end{equation}

Thus, we get a system of differential equations:
\begin{equation}
\begin{cases}
\cfrac{d U^{0 '}}{d \lambda} = U^{1 '} \cfrac{d \alpha}{d \lambda} = U^{1 '} \cfrac{\partial \alpha}{\partial t} \cfrac{d t}{d \lambda} + U^{1 '} \cfrac{\partial \alpha}{\partial r} \cfrac{d r}{d \lambda}
\\
\cfrac{d U^{1 '}}{d \lambda} = U^{0 '} \cfrac{d \alpha}{d \lambda} = U^{0 '} \cfrac{\partial \alpha}{\partial t} \cfrac{d t}{d \lambda} + U^{0 '} \cfrac{\partial \alpha}{\partial r} \cfrac{d r}{d \lambda}
\end{cases}
\end{equation}

Considering $U^{0 '}=\sqrt{g_{00}} U^{0}$ , $U^{1 '}=\sqrt{|g_{11}|} U^{1}$ , we can rewrite it in the form:
\\


\begin{equation}\label{alphatransport}
\begin{cases}
\cfrac{d  U^{0}}{d \lambda} = -\frac{1}{ \sqrt{g_{00}}}  \cfrac{\partial \sqrt{g_{00}}}{\partial t} U^{0} \cfrac{d t}{d \lambda} -\frac{1}{ \sqrt{g_{00}}}  \cfrac{\partial \sqrt{g_{00}}}{\partial r} U^{0} \cfrac{d r}{d \lambda} + \frac{\sqrt{|g_{11}|}}{ \sqrt{g_{00}}}  \cfrac{\partial \alpha}{\partial t}   U^{1} \cfrac{d t}{d \lambda} + \frac{\sqrt{|g_{11}|}}{ \sqrt{g_{00}}}  \cfrac{\partial \alpha}{\partial r} U^{1} \cfrac{d r}{d \lambda}
\\
  \cfrac{d  U^{1}}{d \lambda} = -\frac{1}{\sqrt{|g_{11}|}} \cfrac{\partial \sqrt{|g_{11}|} }{\partial t} U^{1}  \cfrac{d t }{d \lambda} - \frac{1}{\sqrt{|g_{11}|}} \cfrac{\partial \sqrt{|g_{11}|} }{\partial r} U^{1}  \cfrac{d r }{d \lambda} + \frac{\sqrt{g_{00}}}{\sqrt{|g_{11}|}}  \cfrac{\partial \alpha}{\partial t}   U^{0} \cfrac{d t}{d \lambda} + \frac{\sqrt{g_{00}}}{\sqrt{|g_{11}|}} \cfrac{\partial \alpha}{\partial r}  U^{0} \cfrac{d r}{d \lambda}
\end{cases}
\end{equation}

This system of equations coincides with the system \eqref{transport}, which, for spherical symmetry, simplifies and has the form:
\begin{equation}
\begin{cases}
\cfrac{d  U^{0}}{d \lambda} = -\Gamma^0 {}_{0 0} U^{0} \cfrac{d t}{d \lambda} -\Gamma^0 {}_{0 1} U^{0} \cfrac{d r}{d \lambda} - \Gamma^0 {}_{1 0} U^{1} \cfrac{d t}{d \lambda} - \Gamma^0 {}_{1 1} U^{1} \cfrac{d r}{d \lambda}
\\
  \cfrac{d  U^{1}}{d \lambda} = -\Gamma^1 {}_{1 0} U^{1}  \cfrac{d t }{d \lambda} - \Gamma^1 {}_{1 1} U^{1}  \cfrac{d r }{d \lambda} - \Gamma^1 {}_{00} U^{0} \cfrac{d t}{d \lambda} - \Gamma^1 {}_{0 1} U^{0} \cfrac{d r}{d \lambda}
\end{cases}
\end{equation}

Equating the same terms with $ U^\mu \frac{dx^\nu}{d \lambda}$, we get the following two equations:
\begin{equation} \label{alphapartials}
    \begin{cases}
    \frac{\sqrt{|g_{11}|}}{ \sqrt{g_{00}}}  \cfrac{\partial \alpha}{\partial t} =  - \Gamma^0 {}_{1 0} 
    \\
    \frac{\sqrt{|g_{11}|}}{ \sqrt{g_{00}}}  \cfrac{\partial \alpha}{\partial r} =  - \Gamma^0 {}_{1 1} 
    \end{cases}
\end{equation}
other equations  are equivalent to these two equations (this can be easily checked using the 
fact that connection we use is the metric connection).

Using this equations  one can  integrate $\alpha$ along the curve, find $\alpha (\lambda)$ and then get the corresponding velocity
$$
 V=\tanh (\alpha (\lambda))=  \tanh (\alpha_{*} + \int^{\lambda}_{\lambda_*} \frac{d \alpha}{d \lambda} d \lambda  ) =  \tanh (\operatorname{artanh}(\pm v_{loc}) + \Delta \alpha).
$$

At this point we can argue that
the choice of Weitzenbock connection (zero curvature and non-zero torsion, see details, for example,
in \cite{connections}) used for formulation of
Teleparallel Equivalent of General Relativity \cite{Pereira}
is not good for describing parallel transport.

We remind a reader that unlike the Levi-Civita connections which are determined solely in therms of metrics,
definition of Weitzenbock connection needs additional structure in the form of tetrad field.
We have 
\begin{equation} \label{wbconn1}
\Gamma^\alpha {}_{\beta \gamma} = h_{A} {}^{\alpha}  \partial_{\gamma} h^{A} {}_{\beta}
\end{equation}
where $ h^{A}_{\alpha}$ is a tetrad field:
\begin{equation} \label{tetradmink}
g_{\alpha \beta } = \eta_{A B } h^A {}_{\alpha} h^{B} {}_{\beta} 
\end{equation}

\begin{equation} 
 \eta_{A B } = diag(1,-1,-1,-1)
\end{equation}

Usually the tetrad field $ h^{A} {}_{\alpha}$ is assumed to be diagonal. However, if so, and the metric is
diagonal as well, then it can be easily verified that $\Gamma^0 {}_{10}=\Gamma^0 {}_{11}=\Gamma^1 {}_{00}=\Gamma^1 {}_{01}=0$. It means that $\alpha$ and then the 3-velocity always remains constant during the transport. So, it gives us that the recession velocities of distant galaxies are zero.

From this point we return to Levi-Civita connection and will consider only it.
If the transport is along the curve $ t = const $ 
we need only one equation for $d\alpha/dr$.
In the particular case of synchronous metrics we can use it to get general 
relation between velocities defined by two different methods of the Sec.2.

In synchronous metrics $g_{00}=1$. We denote for brevity $-g_{11}=g_1(t,r)$.

Using \eqref{levicon} we calculate  the Christoffel symbols:
\begin{equation} \label{properchristoffel}
\begin{matrix}
\Gamma^0 {}_{0 1}=0
\\
\Gamma^0 {}_{1 1} =\cfrac{1}{2} \cfrac{\partial g_1 }{\partial t}
\\
\Gamma^1 {}_{0 1} =\cfrac{1}{2 g_1} \cfrac{\partial g_1 }{\partial t}
\\
\Gamma^1 {}_{1 1} =\cfrac{1}{2 g_1} \cfrac{\partial g_1 }{\partial r}
\end{matrix}
\end{equation}

Choosing in  \eqref{properchristoffel} and \eqref{alphapartials} the equations with $-\Gamma^0 {}_{1 1}$  and equating them, we get that $ \cfrac{d \alpha}{dr}= -\cfrac{1}{2 \sqrt{g_1}} \cfrac{\partial g_1 }{\partial t}= -\cfrac{\partial \sqrt{g_1} }{\partial t}$. When  the emitter has bigger radial coordinate r than the observer we have 
\begin{equation}
\alpha (r_{obs})-\alpha (r_{*}) = \Delta \alpha = - \int^{r_{obs}}_{r_*} \cfrac{d \sqrt{g_1} }{d t} dr = - \int^{r_1}_{r_2} \cfrac{d \sqrt{g_1} }{d t} dr = \int^{r_2}_{r_1} \cfrac{d \sqrt{g_1} }{d t}=v_{fl}
\end{equation}

Hence,
\begin{equation}\label{mainformula}
V  =\tanh\left( \operatorname{artanh}(v_{loc})+v_{fl})\right)
\end{equation}
 - it is the velocity of recession from the observer- when the emitter is receding from the center it is receding from the the observer as well. For the particular case of emitter with zero peculiar velocity
we have a very similar expression, relating these two velocities: $V=\tanh (v_{fl})$. This formula was obtained in \cite {Hodorowski} for the Friedman metric. Now we can see that it is valid for any synchronous
spherically symmetric metrics, and that the result can be generalized to non-zero peculiar velocities
of the emitter with the modification of this formula given by (26).

And when  the emitter has the smaller radial coordinate r than the observer
 \begin{equation}
\alpha (r_{obs})-\alpha (r_{*}) = \Delta \alpha = - \int^{r_{obs}}_{r_*} \cfrac{d \sqrt{g_1} }{d t} dr = - \int^{r_2}_{r_1} \cfrac{d \sqrt{g_1} }{d t}=-v_{fl}
\end{equation}

And
\begin{equation}\label{mainformula1}
V  =\tanh\left( \operatorname{artanh}(v_{loc})-v_{fl})\right)
\end{equation}
 
Note that this velocity $V$ is positive if it is directed to bigger values of radial coordinate $r$,
and negative in the opposite case.
  If, however, we want to define  the velocity of recession $V_{recession}$ seen by the observer, looking "inside" in the
direction of emitter, we need to attribute positive sign of the velocity if is directed inward, and negative
sign if it is directed outward. So that
 $V_{recession}=-V=\tanh  \left( v_{fl}-\operatorname{artanh}(v_{loc})\right)$
 - when the emitter is receding from the center of the coordinate system it is approaching the observer.

Note, that if $v_{loc}=0$, we have the same formula in both cases: $V_{recession}=\tanh (v_{fl})$.
 
 Thus, we  have obtained that in any spherically symmetric synchronous metrics the recession velocity defined through the parallel transport is expressed through hyperbolic tangent of the velocity defined as a derivative.

\section{Some particular examples}
In this section we consider certain important metrics and show what the results of the proceeding
section mean for known physical situations.

\subsection{Cosmological metric}

We start with the FRW cosmology. Since cosmological recession velocities are very well known both from
their apologists (see for example, \cite{Davis}) and critics (see, for example, \cite{Hodorowski}),
we mostly concentrate on the properties of transported velocity. It seems that this conception 
is almost totally ignored by adepts of the other proposal, so that, ironically, any critical
considerations of this conception are less presented in the methodological literature than the
critics of "standard" recession velocities.

The FRW metrics has the form
\begin{equation}\label{metricfr}
ds^2= dt^2 - a^2(t)\left(dr^2+R^2_{0} S^2(r/R_0) d\theta^2 + R^2_{0} S^2(r/R_0) \sin^2 \theta d\phi^2\right)
\end{equation}
where $S(r)=(\sin{r}, r, \sinh{r})$ for closed, flat and open models respectively.

The Christoffel symbols are

\begin{equation}
\begin{matrix}
\Gamma^0 {}_{11} =a \dot{a}
\\
\Gamma^0 {}_{22} = a \dot{a} R^2_{0} S^2(r/R_0)
\\
\Gamma^0 {}_{33} = a \dot{a} R^2_{0} S^2(r/R_0) \sin^2 \theta  
\\
\Gamma^1 {}_{01} = \Gamma^1 {}_{10} = \Gamma^2 {}_{02} = \Gamma^2 {}_{20} =\Gamma^3 {}_{03} = \Gamma^3 {}_{30} = \cfrac{\dot{a}}{a}
\\
\Gamma^1 {}_{22} = -R_{0} S(\frac{r}{R_0}) S'(\frac{r}{R_0})
\\
\Gamma^1 {}_{33} = -R_{0} S(\frac{r}{R_0}) S'(\frac{r}{R_0}) \sin^2 \theta
\\
\Gamma^2 {}_{12} = \Gamma^2 {}_{21} =\Gamma^3 {}_{13} = \Gamma^3 {}_{31} = \cfrac{S'(\frac{r}{R_0})}{R_{0} S(\frac{r}{R_0})}
\\
\Gamma^2 {}_{33} = - \cos \theta \sin \theta
\\
\Gamma^3 {}_{23} = \Gamma^3 {}_{32} = \cot \theta

\end{matrix}
\end{equation}

We start with the parallel transport along $t=const$ radial line.
First, consider the transport of 4-velocity of an object in the Hubble flow (no peculiar velocity).
The 4-velocity of an emitter is 

\begin{equation}
U^*=(1~,~0~,~0~,~0)
\end{equation}

The system of equations for the parallel transport has the form
\begin{equation}\label{teq}
\cfrac{dU^\alpha}{dx^1} = -\Gamma^\alpha {}_{\beta 1} U^\beta
\end{equation}

\begin{equation} \label{teqfr}
\begin{cases}
\cfrac{dU^0}{dx^1} = -\Gamma^0 {}_{1 1} U^1
\\
\cfrac{dU^1}{dx^1} = -\Gamma^1 {}_{0 1} U^0
\\
\cfrac{dU^2}{dx^1} = -\Gamma^2 {}_{2 1} U^2
\\
\cfrac{dU^3}{dx^1} = -\Gamma^3 {}_{3 1} U^3
\end{cases}
\end{equation}

Third and forth equations decouple, and with the initial conditions  they are trivially satisfied.
The first and second equation after substitution of the corresponding Christoffel  symbols give
\begin{equation}
\begin{cases}
\cfrac{dU^0}{dx^1} = -a \dot{a} U^1
\\
\cfrac{dU^1}{dx^1} = -\cfrac{\dot{a}}{a} U^0
\end{cases}
\end{equation}

The solution of this system is
\begin{equation}
U=(\cosh [ \frac{\dot{a}}{a} d ]~,~\frac{1}{a} \sinh [ \frac{\dot{a}}{a} d ]~,~0~,~0)=(\cosh [ Hd ]~,~\frac{1}{a} \sinh [ Hd ]~,~0~,~0)
\end{equation}
so that 
\begin{equation}
V=\tanh(Hd)=\tanh v_{fl}
\label{main}
\end{equation}
where we use $d=ar$ to denote the physical distance to the emitter. This result was first obtained
in \cite{Hodorowski}. From this formula we immediately see that the additivity, mentioned in Introduction,
does not hold for transported velocities. If one source is located at comoving coordinate $r_1$, and the second
at $r_2$, then recession velocity of the second object $H d_2$ is equal to recession velocity of the
first object $H d_1$ plus the recession velocity of the second object with respect to the first one $H (d_2-d_1)$. As the transported velocity is connected with recession velocity by the formula (36), the additivity
of transported velocities is absent simply because $\tanh(v)$ is not a linear function. Moreover, 
expressing this function through exponents, it is easy to show that these velocities should satisfy the
Special Relativity rule $v=(v_1+v_2)/(1+v_1 v_2)$ instead of simple Galilean rule $v=v_1+v_2$.

Now we consider an object with non-zero peculiar velocity. The 4-velocity of the object in question is

\begin{equation}
U^*=(\frac{1}{\sqrt{1-v^2_{loc}}} ~,~\frac{1}{a^*}  \frac{v_{loc}}{\sqrt{1-v^2_{loc}}}~,~0~,~0)
\end{equation}

After the parallel transport we get

\begin{equation}
\begin{matrix}
U=(\cosh [ \frac{\dot{a}}{a} d + \operatorname{artanh} (v_{loc}) ]~,~\frac{1}{a} \sinh [ \frac{\dot{a}}{a} d + \operatorname{artanh} (v_{loc})]~,~0~,~0)
\\
=(\cosh [ Hd + \operatorname{artanh} (v_{loc})]~,~\frac{1}{a} \sinh [ Hd + \operatorname{artanh} (v_{loc})]~,~0~,~0).
\end{matrix}
\end{equation}

For corresponding 3-velocity we obtain 
\begin{equation}
V=\tanh[ Hd + \operatorname{artanh} (v_{loc})]=\tanh[ v_{fl} + \operatorname{artanh} (v_{loc})] 
\end{equation}

as it should be.

If we subtract the transported Hubble flow velocity from this result, we should get a naive analog of  peculiar velocity.
 Note, however, that this "peculiar velocity" is not equal to $v_{loc}$, and, moreover,
depends on the distance to the object. Vice versa, a peculiar velocity as an intrinsic property of an
emitter is {\it not} equal to the difference between the transported velocities of the object and   of
the Hubble flow in its location. Again, we should use here the Special Relativity velocity-addition formula.

Now we consider parallel transport along the light cone. As the light equation of motion reads $dt=-adr$,
the system for parallel transport is
\begin{equation}
\begin{cases}
\cfrac{dU^0}{dt} =   \dot{a}  U^1
\\
\cfrac{dU^1}{dt} = -\cfrac{\dot{a}}{a} U^1 + \cfrac{\dot{a}}{a^2} U^0

\end{cases}
\end{equation}

Starting from  \begin{equation}
U^*=(1~,~0~,~0~,~0)
\end{equation}            
we have
\begin{equation}
\begin{cases}
\cfrac{dU^0}{dx^1} = - \Gamma^0 {}_{1 1} U^1
\\
\cfrac{dU^1}{dx^1} = a \Gamma^1 {}_{1 0} U^1 - \Gamma^1 {}_{0 1} U^0
\\
\cfrac{dU^2}{dx^1} = (a \Gamma^2 {}_{2 0} - \Gamma^2 {}_{2 1}) U^2
\\
\cfrac{dU^3}{dx^1} = (a \Gamma^3 {}_{3 0} - \Gamma^3 {}_{3 1}) U^3
\end{cases}
\end{equation}

and using equation of motion for the light, finally get
\begin{equation}
\begin{cases}
\cfrac{dU^0}{da} =   U^1
\\
\cfrac{dU^1}{da} = -\cfrac{1}{a} U^1 + \cfrac{1}{a^2} U^0

\end{cases}
\end{equation}

Solving this system for the object in the Hubble flow

we get
\begin{equation}
U=(\cfrac{a^2+a_*^2}{2 a a_*} ~ , ~ \cfrac{1}{2 a_*} -\cfrac{a_*}{ 2 a^2}   ~,~ 0 ~ ,~ 0  ) = (\cosh [ \ln (\cfrac{a}{a_*})] ~ , ~ \frac{1}{a} \sinh [ \ln (\cfrac{a}{a_*})]   ~,~ 0 ~ ,~ 0  )
\end{equation}

with corresponding 3-velocity

\begin{equation}
    V=\cfrac{-a^2+a_*^2}{a^2+a_*^2} = \tanh[ \ (\cfrac{a}{a_*})]
\end{equation}

This 3-velocity gives the cosmological redshift coinciding with the relativistic Doppler formula
\begin{equation}
    1+z = \sqrt{\cfrac{1+V}{1-V}} = \cfrac{a}{a_*}
\end{equation}

as it should be for this particular definition of velocity.

For completeness we also write down the 3-velocity of an object with peculiar motion
\begin{equation}
    V=\tanh[ \ln (\cfrac{a}{a_*}) + \operatorname{artanh} (v_{loc})]
\end{equation}

which gives the redshift

\begin{equation}
    1+z =\sqrt{\cfrac{1+V}{1-V}}  = \cfrac{a}{a_*} \sqrt{\cfrac{1+v_{loc}}{1-v_{loc}}}.
\end{equation}

\subsection{Spherically symmetric black hole metric}
In this subsection we consider static spherically symmetric black hole metrics.
We start to consider this metric in stationary coordinates. This case is not covered
by the general result of the preceding section (since it is valid only for synchronous coordinate),
however, stationary coordinate system is the most popular one for black hole description,
so we consider it first. As the coordinate system is not synchronous, we have no natural method
to define an analog of the  Hubble flow. On the contrary, there is no problem to define transported 
velocities.

The static spherically symmetric space-time in the stationary coordinate looks like
\begin{equation} \label{bh}
ds^2=f(r) dt^2-\cfrac{1}{f(r)}dr^2-r^2 d\theta^2 - r^2 \sin^2 \theta d\phi^2
\end{equation}

so  non-zero Christoffel  symbols are
\begin{equation}
\begin{matrix}
\Gamma^0 {}_{01} = \Gamma^0 {}_{10} = \cfrac{f'}{2f}
\\
\Gamma^1 {}_{00} = \frac{1}{2} f f'
\\
\Gamma^1 {}_{11} = -\cfrac{f'}{2f}
\\
\Gamma^1 {}_{22} = - r f
\\
\Gamma^1 {}_{33} = - r f \sin^2 \theta 
\\
\Gamma^2 {}_{12} = \Gamma^2 {}_{21} = \Gamma^3 {}_{13} = \Gamma^3 {}_{31} = \cfrac{1}{r}
\\
\Gamma^2 {}_{33} = - \cos \theta \sin \theta
\\
\Gamma^3 {}_{23} = \Gamma^3 {}_{32} = \cot \theta
\end{matrix}
\end{equation}

The equations for parallel transport along $t=const$ are

\begin{equation}
\begin{cases}
\cfrac{dU^0}{df} = -\cfrac{1}{2f} U^0
\\
\cfrac{dU^1}{df} = \cfrac{1}{2f} U^1
\end{cases}
\end{equation}

For an emitter being at rest ($r=const$), the 4-velocity is
\begin{equation}
U^*=(\cfrac{1}{\sqrt{f^*}}~,~0~,~0~,~0)
\end{equation}

If the emitter moves with respect to the stationary frame in radial direction with the local
velocity $v_{loc}$, its 4-velocity is
\begin{equation}
U^*=(\cfrac{1}{\sqrt{f^*}}  \cfrac{1}{\sqrt{1-v^2_{loc}}}~,~\sqrt{f^*} \cfrac{v_{loc}}{\sqrt{1-v^2_{loc}}}~,~0~,~0).
\end{equation}

Solving the system, we get that in both cases 
\begin{equation}
\begin{cases}
\sqrt{f} U^0=\sqrt{f^*} U_{*}^0
\\
\frac{U^1}{\sqrt{f}}  =\frac{U_{*}^1}{\sqrt{f^*}}
\end{cases}
\end{equation}

\begin{equation}
\begin{cases}
U^{0'}= U_{*}^{0'}= const
\\
U^{1'} =U_{*}^{1'}= const
\end{cases}
\end{equation}

which means that the resulting 3-velocity coincides with $v_{loc}$.

This results looks like a trivial one. However, if we consider 
 transport along the light line 
we get a kind of counter-intuitive result
(though not unexpected, as we will see soon). Indeed, as the equation
for the light propagation gives us  $f dt=\pm dr $ (upper sign used if the observer has bigger radial coordinate $r$ than the emitter,
and the lower sign in the opposite situation), and we get the following equations for the parallel transport

\begin{equation}
\begin{cases}
\cfrac{dU^0}{dr} = \mp \cfrac{f'}{2f^{2}} U^1 - \cfrac{f'}{2f} U^0
\\
\cfrac{dU^1}{dr} = \mp \cfrac{1}{2}  f' U^0 + \cfrac{f'}{2f} U^1,

\end{cases}
\end{equation}

or 

\begin{equation}
\begin{cases}
\cfrac{dU^0}{df} = \mp \cfrac{1}{2f^{2}} U^1 - \cfrac{1}{2f} U^0
\\
\cfrac{dU^1}{df} = \mp \cfrac{1}{2}  U^0 + \cfrac{1}{2f} U^1.

\end{cases}
\end{equation}

Now, the solution for the 4-velocity of an emitter "at rest" 
is
\begin{equation}
U =  (\cfrac{1}{\sqrt{f}} \cosh [ \ln (\sqrt{\frac{f}{f^*}})] ~ , ~ \mp \sqrt{f}~ \sinh [ \ln (\sqrt{\frac{f}{f^*}})]   ~,~ 0 ~ ,~ 0  ) 
\end{equation}

which gives us the 3-velocity
\begin{equation}
    V= \mp  \tanh [ \ln (\sqrt{\frac{f}{f^*}})]
\end{equation}

Velocity of recession from the observer is $V_{recession}=\mp V=   \tanh [ \ln (\sqrt{\frac{f}{f^*}})]$
(where we should use lower sign if the emitter has bigger radial coordinate $r$ than the observer,
and the upper sign in the opposite situation).

We see that after adopting the procedure of parallel transport along light-line curve we get a nonzero
velocity of an object being at rest with respect to stationary coordinate system when observed
by an observer which is also at rest. Obviously, the proper distance between rest observer and 
rest emitter in stationary coordinates does not change with time. However, the non-zero transported
3-velocity in this situation is not unexpected because, as we note above, this velocity coincides
with the velocity corresponding to the observed redshift if interpreted as a standard relativistic
Doppler effect. Indeed, it is possible to show that the velocity found induces the Doppler shift 
which is equal to

\begin{equation}
    1+z = \sqrt{\frac{1 \mp V}{1 \pm V}} = \sqrt{\frac{1 + V_{recession}}{1 - V_{recession}}} = \sqrt{\frac{f}{f^*}}
\end{equation}
(with the same sign convention as above),
so the usual gravitational redshift is interpreted as a Doppler shift.

The situation when two rest particles in stationary coordinates (so that nothing  changes in
time at all!) have non-zero mutual velocity is a good illustration of the point mentioned in Introduction-
when velocity vector undergoes a parallel transport it is no longer a velocity vector of some
physical object, but rather can be considered as an abstract mathematical vector.

For completeness we note that for the particle moving in radial direction 

\begin{equation}
    U^*=(\cfrac{1}{\sqrt{f^*}} \cfrac{1}{\sqrt{1-v^2_{loc}}} ~,~\sqrt{f^*} \cfrac{ v_{loc}}{\sqrt{1-v^2_{loc}}}~,~0~,~0)
\end{equation}

and the resulting 3-velocity is
\begin{equation}
    V  =\tanh[\operatorname{artanh} (v_{loc}) \mp  \ln  (\sqrt{\frac{f}{f^*}} )]
\end{equation}
with the expected form of redshift
\begin{equation}
    1+z =\sqrt{\frac{1 + V_{recession}}{1 - V_{recession}}} = \sqrt{\frac{1 \mp V}{1 \pm V}} = \sqrt{\frac{f}{f^*}} \sqrt{\frac{1 \mp v_{loc}}{1 \pm v_{loc}}}
\end{equation}

We now return to synchronous coordinate, so that we will consider particles free falling into 
a black hole. The coordinates associates with such particles generalize well known Lemaitre
coordinates for the Schwarzschild space-time and have the line element

\begin{equation} \label{metriclem}
ds^2=d \tau^2-(1-f)d \rho^2-r^2 d\theta^2 - r^2 \sin^2 \theta d\phi^2
\end{equation}

It can be derived from the static coordinate after
making the coordinate transformations 

\begin{equation}
\rho =t+\int \frac{dr}{f\sqrt{1-f}}\text{,}  \label{rof}
\end{equation}%
\begin{equation}
\tau =t+\int \frac{dr}{f}\sqrt{1-f}  \label{tauf}
\end{equation}%

 The role of $r$ in the Lemaitre metrics becomes clear if we calculate the proper distance
$d=\int^{\rho_2}_{\rho_1} d \rho \sqrt{|g_{\rho \rho}|} = \int^{\rho_2}_{\rho_1} d \rho \sqrt{1-f}=
r_2-r_1$ (here we assume that $r_2>r_1$).

So that,for the velocity of a "free-fall flow" -- an analog of Hubble flow in which particles
with $\rho=const$ are participating -- we can try to take $v=dr/d \tau=-\sqrt{1-f}$. 
This velocity $v$ reaches the speed of light at the horizon and continues to grow with decreasing $r$.
In a Schwarzschild  black hole it diverges at singularity.


However, in contrast to FRW cosmology, now the position of an observer is important since there is no spatial homogeneity. If the emitter
is located at $r_1$ and the observer at $r_2$ than their relative velocity is equal to $|v(r_1)-v(r_2)|$. 
It is the relative velocity which an observer should attribute to a distant point to measure how the proper
distance between him/her and the emitter changes with time.
Properties of relative velocity require a special attention to avoid possible confusions. We start
with the case of zero peculiar velocities -- let both observer and emitter have constant $\rho$.
First of all, if an emitter crossed a horizon, its velocity with respect to any observer located
at final $r$ is subluminal since $v(r_g)=1$ and $v(r_2)$ is nonzero and positive. Remember,
on the contrary, that any horizon crossing object has the velocity {\it with respect
to stationary coordinate system} equal to speed of light independently of velocity of an observer
from outside. The relative velocity $v(r_1)-v(r_2)$ reaches the speed of light somewhere inside the horizon
depending on the position of the observer $r_2$. On the other side, if we include peculiar 
velocities, the overall velocity can be superluminal even for both observer and emitter being
outside the horizon. For example, the light beam at the point $r_1$ propagating inward 
from the viewpoint of an observer at $r_2$ has the velocity $c+v(r_1)-v(r_2)$ which
is superluminal for any observer with $r_2>r_1$.

As for the transported velocities, the calculation is straightforward, and the results match with the general
results obtained in the previous section.  After substitution the appropriate Christoffel  symbols we get 
the relevant equations for the transfer along $\tau=const$ in the form

\begin{equation}
\begin{cases}
\cfrac{dU^0}{d\rho} = -\cfrac{1}{2} \cfrac{d f}{d \rho}  U^1
\\
\cfrac{dU^1}{d\rho} =\left(\cfrac{d f}{d \rho}\right) \cfrac{U^1-U^0}{2(1-f)}
\end{cases} 
\end{equation}

For the emitter at rest in Lemaitre frame   $U^*=(1~,~0~,~0~,~0)$  the solution is 
\begin{equation}
\begin{cases}
U^0=\cosh [ \sqrt{1-f} - \sqrt{1-f^*} ]
\\
U^1=\cfrac{\sinh [ \sqrt{1-f} - \sqrt{1-f^*} ]}{\sqrt{1-f}}
\end{cases}
\end{equation}

which  gives us the 3-velocity   $V=\tanh[ \sqrt{1-f} - \sqrt{1-f^*} ] = \tanh[ \mp v_{fl}  ]
$  in accordance with  \eqref{mainformula}.
Similar way, the 4-velocity of an emitter with non-zero peculiar velocity is
\begin{equation}
U^*=(\frac{1}{\sqrt{1-v^2_{loc}}}~,~ \frac{1}{\sqrt{1-f^*}} \frac{v_{loc}}{\sqrt{1-v^2_{loc}}}~,~0~,~0)
\end{equation}

and the solution of the system   gives  
\begin{equation}
\begin{cases}
U^0=\cosh [ \sqrt{1-f} - \sqrt{1-f^*} + \operatorname{artanh} (v_{loc})]
\\
U^1=\cfrac{\sinh [ \sqrt{1-f} - \sqrt{1-f^*} + \operatorname{artanh} (v_{loc})]}{\sqrt{1-f}}
\end{cases}
\end{equation}

and $V=\tanh [ \sqrt{1-f} - \sqrt{1-f^*} + \operatorname{artanh} (v_{loc})]=\tanh[\mp v_{fl}  + \operatorname{artanh} (v_{loc})]$ as it should be.

For the transport along the light line we should use that   

$\tau =\tau(\rho)$, and    $d\tau=\pm \sqrt{1-f}d\rho$, and as $f(\tau , \rho ) = f(\tau (\rho) , \rho ) $ then $\cfrac{df}{ d \rho} = \cfrac{\partial f}{ \partial \rho} \pm \cfrac{\partial f}{ \partial \tau } \cfrac{\partial \tau}{ \partial \rho}  = \cfrac{\partial f}{ \partial \rho} (1 \mp \sqrt{1-f}) $  (where the upper sign is for the case when the observer has bigger radial coordinate).

Now the system for the parallel transport takes the form

\begin{equation}
\begin{cases}
(1 \mp \sqrt{1-f}) \cfrac{dU^0}{d \rho} =  -\cfrac{1}{2} \cfrac{d f}{d \rho} U^1
\\
(1 \mp \sqrt{1-f}) \cfrac{dU^1}{d \rho}  =  \cfrac{(1 \mp \sqrt{1-f})  U^1 - U^0 }{2(1-f)}  \left( \cfrac{d f}{d \rho} \right),

\end{cases}
\end{equation}

or 

\begin{equation}
\begin{cases}
(1 \mp \sqrt{1-f}) \cfrac{dU^0}{d f} =  -\cfrac{1}{2} U^1
\\
(1 \mp \sqrt{1-f}) \cfrac{dU^1}{d f}  =  \cfrac{(1 \mp \sqrt{1-f})  U^1 - U^0 }{2(1-f)}  

\end{cases}
\end{equation}

The resulting 4-velocity for the emitter at rest is
\begin{equation}
\begin{cases}

U^0=\cosh  \left[   \ln   \left( \cfrac{1 \mp \sqrt{1-f}}{1 \mp \sqrt{1-f^*}}  \right) \right]
\\
U^1=\cfrac{1}{\sqrt{1-f}} \sinh   \left[ \mp    \ln  \left( \cfrac{1 \mp \sqrt{1-f}}{1 \mp \sqrt{1-f^*}}  \right) \right]

\end{cases}
\end{equation}

and the recession velocity

\begin{equation}
 V_{recession}=\pm V  =\tanh\left[   \ln   \left( \cfrac{1 \mp \sqrt{1-f}}{1 \mp \sqrt{1-f^*}}  \right) \right]
\end{equation}

The Doppler shift corresponding to this velocity is
\begin{equation}
    1+z =\sqrt{\cfrac{1 \mp V}{1 \pm V} } = \cfrac{1 \mp \sqrt{1-f}}{1 \mp \sqrt{1-f^*}}
\end{equation}

Not the difference between redshift expressed through the function $f$
for observer and emitter at rest with respect to the
stationary coordinate system (60) and with respect to the free-falling coordinate system (75).
For completeness, nonzero peculiar velocity leads to resulting 4-velocity

\begin{equation}
\begin{cases}

U^0=\cosh  \left[ \mp    \ln   \left( \cfrac{1 \mp \sqrt{1-f}}{1 \mp \sqrt{1-f^*}}  \right) +\operatorname{artanh}( v_{loc} ) \right ]
\\
U^1=\cfrac{1}{\sqrt{1-f}} \sinh   \left[  \mp    \ln  \left( \cfrac{1 \mp \sqrt{1-f}}{1 \mp \sqrt{1-f^*}}\right) + \operatorname{artanh}( v_{loc} )  \right]

\end{cases}
\end{equation}

3-velocity
\begin{equation}
    V=  \tanh  \left[  \mp  \ln  \left( \cfrac{1 \mp \sqrt{1-f}}{1 \mp \sqrt{1-f^*}}  \right) + \operatorname{artanh}( v_{loc} )  \right]
\end{equation}

and corresponding redshift

\begin{equation}
    1+z =\sqrt{\cfrac{1 \mp V}{1 \pm V}}   = \cfrac{1 \mp \sqrt{1-f}}{1 \mp \sqrt{1-f^*}} \sqrt{\cfrac{1  \mp  v_{loc}}{1 \pm v_{loc}}}
\end{equation}

\section{Conclusions}

In the present paper we have considered two different definitions of velocities of remote objects in
General Relativity. Since both are mathematically correct (if the foliation of the space-time in
question by hypersurfaces of constant time is given), it is possible to use them in appropriate
situations without any problems. However, it is necessary to remember that properties of these
velocities may look strange in comparison with  "usual" velocities in classical physics and even
in Special Relativity, and using intuition instead of calculations might be dangerous.

The main feature which distinguishes the GR situation from Special Relativity in the first 
case considered in the present paper -- the velocity defined as the derivative of proper distance
to the object with respect to the proper time of the observer -- is the possibility for the velocity to be 
superluminal.  This fact have been discussed many times in the cosmological situation, and 
superluminal cosmological recession velocities now are accepted in scientific community. In the present
paper we show that this situation is not strictly connected with cosmology, but appears in any
synchronous coordinate system. FRW frame, being synchronous, is the most natural for homogeneous and isotropic cosmology,
so not surprising that this property is well known in a cosmological situation. However, the same picture
appears in black hole space-times if we use the synchronous Lemaitre frame to define $t=const$
hypersurfaces. It is also curious that not only a free fall inside the event horizon can be
superluminal, but even the motion of a particle, located outside the horizon, but having non-zero 
peculiar velocity with respect to
the Lemaitre frame and directed inward can be superluminal as well. The same situation
in cosmology (when the resulting velocity of a particle which has big enough peculiar velocity directed
outward can exceed speed of light even if the particle is located within the Hubble sphere)
is also possible, though it seems to look 
less counter-intuitive (possibly because the Hubble sphere is obviously observer dependent in
contrast to the black hole horizon). The other "strange" feature of the velocity in question is that it can be 
subluminal for a particle crossing event horizon, and reach the speed of light somewhere
inside the horizon (depending on the position of the observer).

The other definition of velocity discussed in the present paper -- the velocity defined by parallel
transport of the initial 4-velocity of the emitter using the Levi-Civita connection -- is free from
such superluminal properties by definition. Since the result of parallel transport depends upon the path,
the velocities defined by the transport along the line of constant time and along the light geodesic
will be different. The former case matches our intuition about the velocity "now". We show that 
it is connected with the velocity defined as a derivative by a simple formula. As for the latter case, the velocity
defined this way allows us to interpret any redshift in the presence of gravity as a Doppler kinematic
shift. This can be counter-intuitive, for example, in stationary black hole metrics, where stationary emitter has
non-zero "velocity" with respect to stationary observer.

We hope that our treatment of some "strange" features of the considered velocities will help 
to use these both definition correctly in appropriate physical situations.

\section*{Acknowledgements}
The work was supported by the Program "Leading  Science School MSU (Physics of Stars, Relativistic Compact Objects and Galaxies)". Authors are grateful to Sergey Popov for discussions.

\end{document}